\documentclass[journal,10pt,twocolumn]{IEEEtran}
\usepackage{xmpaper,lzqsymbol}
\usepackage{amssymb,amsmath,latexsym}
\usepackage{graphicx,subfigure}
\usepackage{color,cite,times}
\usepackage{epstopdf}
\usepackage{multirow}
\usepackage{array}

\usepackage{amsmath}
\usepackage{bm}
\usepackage{algorithm}
\usepackage{algorithmic}
\usepackage{multicol}
\usepackage{pdfpages}
\usepackage[T1]{fontenc}

\usepackage{booktabs}
\usepackage{bbm} 
\usepackage{subfigure}

\begin{document}
\title{Reconfigurable Intelligent Surface Assisted UAV Communication: Joint Trajectory Design and Passive Beamforming}
\author{Sixian Li, Bin Duo, Xiaojun~Yuan,~\IEEEmembership{Senior~Member,~IEEE}, Ying-Chang~Liang,~\IEEEmembership{Fellow,~IEEE}, and Marco~Di~Renzo,~\IEEEmembership{Fellow,~IEEE}

\thanks{S.~Li, B.~Duo, X.~Yuan, and Y.-C.~Liang are with  the National Laboratory of Science and Technology on Communications, the University of Electronic Science and Technology of China, Chengdu 611731,
China (e-mail: sxli@std.uestc.edu.cn; duobin@cdut.edu.cn; xjyuan@uestc.edu.cn; liangyc@ieee.org). M.~Di~Renzo is with Paris-Saclay University (L2S - CNRS, CentraleSupelec, University Paris Sud), Paris, France (e-mail: marco.direnzo@l2s.centralesupelec.fr).
}
}

\maketitle

\begin{abstract}
    Thanks to the line-of-sight (LoS) transmission and flexibility, unmanned aerial vehicles (UAVs) effectively improve the throughput of wireless networks. Nevertheless, the LoS links are prone to severe deterioration by complex propagation environments, especially in urban areas. Reconfigurable intelligent surfaces (RISs), as a promising technique, can significantly improve the propagation environment and enhance communication quality by intelligently reflecting the received signals. Motivated by this, the joint UAV trajectory and RIS's passive beamforming design for a novel RIS-assisted UAV communication system is investigated to maximize the average achievable rate in this letter. To tackle the formulated non-convex problem, we divide it into two subproblems, namely, passive beamforming and trajectory optimization. We first derive a closed-form phase-shift solution for any given UAV trajectory to achieve the phase alignment of the received signals from different transmission paths. Then, with the optimal phase-shift solution, we obtain a suboptimal trajectory solution by using the successive convex approximation (SCA) method. Numerical results demonstrate that the proposed algorithm can considerably improve the average achievable rate of the system.
\end{abstract}
\section{Introduction}
    With the rapid development of the fifth-generation $(5\rm{G})$ wireless networks, unmanned aerial vehicles (UAVs) are playing an increasingly significant role in improving spectral efficiency \cite{Gupta2016}. Owing to their high mobility, line-of-sight (LoS) transmission, and low cost, UAVs have been widely used in various scenarios to enhance communication quality via jointly optimizing UAV trajectory and communication resource allocation, including the common throughput, average secrecy rate, and energy efficiency maximization, etc. \cite{duo2019, zhang2019}
    
    To improve the propagation environment and enhance communication quality, reconfigurable intelligent surface (RIS) has attracted extensive attention \cite{huang2019, liang2019}. Generally, a RIS is comprised of abundant reconfigurable reflecting elements that are energy-efficient and cost-effective. Each element in the RIS can reflect the incident signal by inducing a manageable phase shift. The phase shifts of all the elements can be jointly adjusted to achieve the phase alignment of the signals from different transmission paths at a desired receiver, also known as passive beamforming, so as to increase the signal energy and improve the achievable rate \cite{wu2018intelligent}. It is worth noting that unlike the conventional amplify-and-forward relay, the RIS reflects the arrived signals by its passive elements, which results in low energy consumption for the RIS. Besides, the RIS has no transmitter module and the elements of the RIS are low-cost \cite{ntontin2019}, which reduces the implementation cost of the RIS. Another advantage is that the RIS can easily work in full-duplex mode without complex interference cancellation techniques.

    \begin{figure}[t]
      \centering
      \includegraphics[width=2.5in]{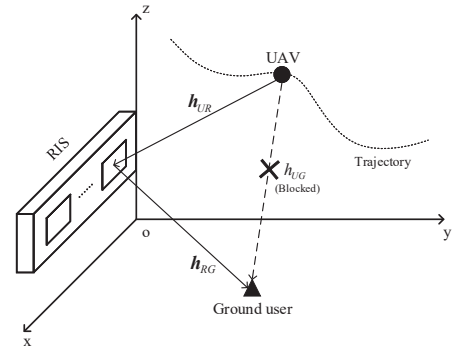}
      \caption{A RIS-assisted UAV communication system.}
      \label{figure_1}
    \end{figure}
    Due to the complex urban environment, the LoS link between the UAV and the ground user may be blocked, which severely degrades the channel quality. Considering the promising RIS technique, we propose a RIS-assisted UAV communication system, where a mobile UAV communicates with a ground user along its planned trajectory, and its transmitted signal is reflected to the user via the RIS. Our goal is to maximize the average achievable rate by jointly designing UAV trajectory and passive beamforming subject to practical UAV mobility and the RIS's phase-shift constraints. To address the non-convexity of the considered problem, we first align the user's received signals from the UAV and the RIS for maximizing the received signal power. Then, a closed-form phase-shift solution can be obtained, and thus the formulated problem is reduced to the optimization of the UAV trajectory. By applying the successive convex approximation (SCA) technique, we finally obtain a locally optimal solution for the joint design problem. Simulation results demonstrate that the proposed algorithm can significantly increase the average achievable rate, as compared to benchmark algorithms.
    
    \section{System Model and Problem Formulation}
    \subsection{System Model}
    In this paper, we consider a downlink transmission system consisting of a rotary-wing UAV\footnote{We assume that the UAV has no effect of UAV jittering and wind speed uncertainty, i.e., the UAV has a perfectly stable flight  \cite{jiang2016, feng2019}. In addition, when the rotary-wing UAV moves, the variations of its Euler angles are relatively small compared with the fixed-wing UAV. Thus, when the UAV flies higher than the RIS, the probability of the LoS path to be blocked from the UAV to the RIS is very small. Hence, the rotational effects can be ignored in this letter.}, a ground user, and a RIS on a building. This system can be extended to more general scenarios, e.g., in a crowded area for grand events, where the surrounding base stations are overloaded with heavy communication traffics. UAVs and RISs are deployed for traffic offloading, so as to enhance the communication quality of ground users. As shown in Fig.~\ref{figure_1}, all communication nodes are placed in the three dimensional (3D) Cartesian coordinate system. The ground user's horizontal coordinate is denoted by $\mathbf{w}_G=[x_G, y_G]^T$. The UAV flies at a fixed altitude denoted by $z_U$ for a finite time span \emph{T}. For tractability, \emph{T} is divided into \emph{N} time slots, i.e., $T = N\delta_t$, where $\delta_t$ is the slot length. As a result, the UAV's horizontal trajectory can be approximated by the sequence $\mathbf{q}[n]=[x[n], y[n]]^T, n \in \mathcal{N}=\{1,\cdots, N \}$, which meets the following mobility constraints:
    \begin{subequations} \label{mobility.1}
    \begin{align}
        & ||\mathbf{q}[n+1]-\mathbf{q}[n]||^{2} \leq D^2,n=1,\cdots,N-1, \label{mobility.1.a}\\
        & ||\mathbf{q}[N]-\mathbf{q}_F||^{2} \leq D^2, \mathbf{q}[1] = \mathbf{q}_0, \label{mobility.1.b}
    \end{align}
    \end{subequations}
    where $\mathbf{q}_0$ and $\mathbf{q}_F$ denote UAV's initial and final horizontal locations, respectively, $D=v_{\max}\delta_t$ is the maximum distance that the UAV can horizontally move within a single time slot, and $v_{\max}$ is the maximum speed of the UAV.
    
    We assume that the UAV and the user are equipped with a single omni-directional antenna, while the RIS is equipped with a uniform linear array (ULA) of $M$ reflecting elements and a controller intelligently adjusting the phase shift of each element. The RIS is located in the $x$-$z$ plane and parallels to the $x$-axis. Let $\boldsymbol{\Theta}[n]=\diag\{ e^{j\theta_1[n]},  e^{j\theta_2[n]},\cdots, e^{j\theta_M[n]}\}$ be the diagonal phase-shift matrix for the RIS in the $n$th time slot, where $\theta_i[n] \in \left[0,2\pi\right), i \in \mathcal{M}=\{1,\cdots, M\}$, is the phase shift of the $i$th reflecting element in time slot $n$, and the phase shifts $\{\theta_i[n]\}$ are assumed to be continuously controllable. Furthermore, the first element of the RIS is regarded as the reference point whose altitude and horizontal coordinates are denoted by $z_R$ and $\mathbf{w}_R=[x_R, y_R]^T$, respectively. Therefore, the distance between the RIS and a certain communication node can be approximated by that between the reference point and the corresponding node.
    
    Since UAVs usually fly at high altitudes and RISs are commonly placed on the facade of a building, the link from the UAV to the RIS (U-R link) is assumed to be a LoS channel. Even if the LoS link from the UAV to the ground user (U-G link) is blocked, there still exist extensive scatters. Thus, we assume the Rayleigh fading channel model for the U-G link. Due to the additional LoS path, the link from the RIS to the ground user (R-G link) can be modeled by a Rician fading channel. Specially, the U-R and R-G links are collectively called the U-R-G link. For clarity, unlike a uniform rectangular array (URA) at the IRS, which can be regarded as a specular reflector \cite{DBLP:journals/corr/abs-1906-09490}, we utilize a ULA at the RIS in this letter. Therefore, the subsequent channel modeling are characterized as the product channels \cite{DBLP:journals/corr/abs-1906-09490}. The channel gain of the U-G link in the $n$th time slot, denoted by $h_{U\!G}[n]$, can be expressed as 
    \begin{equation} \label{gain.1}
        h_{U\!G}[n]= \sqrt{\rho d_{U\!G}^{-\kappa}[n]}~\tilde{h},
    \end{equation}
     where $\rho$ is the path loss at the reference distance $D_0=1$ m \cite{wu2018intelligent}, $\kappa$ is the corresponding path loss exponent related to the U-G link, $d_{U\!G}[n]= \sqrt{z_U^2+||\mathbf{q}[n]-\mathbf{w}_G||^2}$ denotes the distance between the UAV and the ground user in the $n$th time slot, and $\tilde{h}$ represents the random scattering component modeled by a zero-mean and unit-variance circularly symmetric complex Gaussian (CSCG) random variable.

    The channel gain of the U-R link in the $n$th time slot, denoted by $\boldsymbol{h}_{UR}[n] \in \mathbb{C}^{M\times 1}$, is given by
    \begin{equation} \label{gain.2}
        \boldsymbol{h}_{U\!R}[n]\!=\!\underbrace{\sqrt{\rho d_{U\!R}^{-2}[n]}}_{\text{path loss}}\!\underbrace{\left[1, e^{-\!j\!\frac{2\pi}{\lambda}d\phi_{U\!R}[n]}, ... , e^{ -\!j\!\frac{2\pi}{\lambda}\!(M\!-\!1)d\phi_{U\!R}[n]}\right]^T}_{\text{array response}},
    \end{equation}
    where $d_{U\!R}[n]=\sqrt{(z_U-z_R)^2+||\mathbf{q}[n]-\mathbf{w}_R||^2}$, the right-most term in \eqref{gain.2} is the array response of an $M$-element ULA \cite{Tse2009Fundamentals}, $\phi_{U\!R}[n]=\frac{x_R-x[n]}{d_{U\!R}[n]}$ represents the cosine of the angle of arrival (AoA) of the signal from the UAV to the ULA at the RIS in the $n$th time slot, $d$ is the antenna separation, and $\lambda$ is the carrier wavelength.

    Similarly, the channel gain of the R-G link, denoted by $\boldsymbol{h}_{RG}\in \mathbb{C}^{M\times 1}$, can be expressed as
    \begin{equation} \label{gain.3}
        \boldsymbol{h}_{RG}=\underbrace{\sqrt{\rho d_{RG}^{-\alpha}}}_{\text{path loss}}\underbrace{\left(\sqrt{\frac{\beta}{1+\beta}}\boldsymbol{h}_{RG}^{\rm{LoS}}+\sqrt{\frac{1}{\beta+1}}\boldsymbol{h}_{RG}^{\rm{NLoS}}\right)}_{\text{array response \& small-scale fading}},
    \end{equation}
    where $d_{RG}=\sqrt{z_I^2+||\mathbf{w}_I-\mathbf{w}_G||^2}$, the summed terms in \eqref{gain.3} include the deterministic LoS component $\boldsymbol{h}_{RG}^{\rm{LoS}}= \left[1, e^{-j\frac{2\pi}{\lambda}d\phi_{RG}}, ... , e^{-j\frac{2\pi}{\lambda}(M-1)d\phi_{RG}}\right]^T \in \mathbb{C}^{M\times 1}$ and the non-LoS (NLoS) component $\boldsymbol{h}_{RG}^{\rm{NLoS}} \in \mathbb{C}^{M\times 1}$ with the variables independently drawn from the CSCG distribution with zero mean and unit variance, $\phi_{RG}=\frac{x_G-x_R}{d_{RG}}$ is the cosine of  the angle of departure (AoD) of the signal from the ULA at the RIS to the user, $\beta$ is the Rician factor, and $\alpha$ is the path loss exponent related to the R-G link. To facilitate the subsequent discussions, the complex vector $\boldsymbol{h}_{RG}$ can also be expressed as below,
    \begin{equation}\label{trans.1}
        \boldsymbol{h}_{RG} = \left[|h_{RG,1}| e^{j\omega_1}, |h_{RG,2}| e^{j\omega_2}, \cdots, |h_{RG,M}| e^{j\omega_M}\right]^T,
    \end{equation}
    where $|h_{RG,i}|$ and $\omega_i\in[0,2\pi)$ are the magnitude and phase angle of the $i$th element of the complex vector $\boldsymbol{h}_{RG}$, respectively. In this letter, we assume that the channel state information (CSI) can be obtained based on existing channel estimation techniques for RIS assisted channels, such as in \cite{he2019,mishra2019}.

    With \eqref{gain.1}-\eqref{gain.3}, the SNR of the ground user in the $n$th time slot can be written as
    \begin{equation} \label{SNR.1}
        \gamma_{U\!G}[n]=\frac{P{\left| h_{U\!G}[n] + \boldsymbol{h}_{RG}^H \boldsymbol{\Theta}[n] \boldsymbol{h}_{U\!R}[n]\right|^2}}{\sigma^2},
    \end{equation}
    where $P$ is the fixed transmit power of the UAV, and $\sigma^2$ is the noise variance. Thus, the achievable rate in bits/second/Hertz (bps/Hz) of the ground user in the $n$th time slot is given by
    \begin{equation} \label{rate.1}
        R_{U\!G}[n]=\log_2(1+\gamma_{U\!G}[n]).
    \end{equation}
    Accordingly, the average achievable rate over $N$ time slots can be expressed as
    \begin{equation} \label{rate.2}
         \Bar{R}=\frac{1}{N} \sum\limits_{n=1}^N R_{U\!G}[n].
    \end{equation}
\subsection{Problem Formulation}
    In this letter, our objective is to maximize the average achievable rate $\Bar{R}$ by jointly optimizing the UAV's trajectory $\mathbf{Q}\triangleq \{\mathbf{q}[n], n\in\mathcal{N}\}$ and the phase-shift matirx $\boldsymbol{\Phi}\triangleq \{\boldsymbol{\Theta}[n], n\in\mathcal{N}\}$ of the RIS over the entire $N$ time slots, subject to UAV's mobility and RIS's phase-shift constraints. Thus, the problem can be formulated as
    \begin{subequations}\label{optimal.1}
    \begin{align}
         & \max\limits_{\mathbf{Q}, \boldsymbol{\Phi}} \quad\Bar{R}\label{optimal.1.a}\\
         & ~~\textrm{s.t.} \quad ~0 \leq \theta_i[n] < 2\pi, \forall n,i, \label{optimal.1.b}\\
         & ~~ \quad ~~~~ (1). \nonumber
     \end{align}
    \end{subequations}
    Although the constraints in \eqref{mobility.1} and \eqref{optimal.1.b} are convex, it is still difficult to solve the problem in \eqref{optimal.1} optimally due to its non-convex objective function with respect to $\mathbf{Q}$ and $\boldsymbol{\Phi}$. In Section III, we propose an efficient algorithm to obtain a suboptimal solution to problem \eqref{optimal.1}.
\section{Proposed Algorithm}
    In this section, we divide problem \eqref{optimal.1} into two subproblems, i.e., the passive beamforming and the UAV trajectory optimization, respectively. For the first subproblem,  we align the phases of the received signals from the U-G and U-R-G links at the user to maximize the received signal energy. Then, a closed-form phase-shift solution for any given UAV trajectory can be obtained. Consequently, problem \eqref{optimal.1} is transformed into the UAV trajectory optimization problem. Finally, a locally optimal trajectory solution to the second subproblem can be obtained by the SCA method.
    
    \subsection{Optimal $\boldsymbol{\Phi}$ for Given $\mathbf{Q}$}
    We first consider the optimization of $\boldsymbol{\Phi}$ for any given $\mathbf{Q}$. 
    With \eqref{trans.1}, $\boldsymbol{h}_{RG}^H \boldsymbol{\Theta}[n] \boldsymbol{h}_{UR}[n]$ can be written as
    \begin{equation} \label{gain.4}
        \boldsymbol{h}_{RG}^H \boldsymbol{\Theta}[n] \boldsymbol{h}_{U\!R}[n]\!=\!\frac{\sqrt{\rho}\sum\limits_{i=1}^M |h_{RG,i}| e^{j(\theta_i[n]-\omega_i-\frac{2\pi}{\lambda}d(i-1)\phi_{U\!R}[n])}}{d_{U\!R}[n]}.
    \end{equation}
    If the signals from different paths are combined coherently at the user, the coherent signal construction can maximize the received signal power, thereby maximizing the achievable rate. Thus, we set $\theta_1[n]-\omega_1=\theta_2[n]-\omega_2-\frac{2\pi}{\lambda}d \phi_{U\!R}[n]=\cdots=\theta_M[n]-\omega_M-\frac{2\pi}{\lambda}d(M-1)\phi_{U\!R}[n]=\arg(\tilde{h})$, or equivalently,
    \begin{equation}\label{Phi.1}
        \theta_i[n] = \arg(\tilde{h}) + \omega_i + \frac{2\pi}{\lambda}d(i-1)\phi_{U\!R}[n], \forall n,i,
    \end{equation}
    which means that we can achieve the phase alignment of the signals at the user for any given UAV trajectory. As such, $\boldsymbol{h}_{RG}^H \boldsymbol{\Theta}[n] \boldsymbol{h}_{UR}[n]$ can be rewritten as
    \begin{equation}
        \boldsymbol{h}_{RG}^H \boldsymbol{\Theta}[n] \boldsymbol{h}_{U\!R}[n]=\frac{e^{j\!\arg(\tilde{h})}\sqrt{\rho}\sum\limits_{i=1}^M |h_{RG,i}|}{d_{U\!R}[n]}. 
    \end{equation}
    Therefore, problem \eqref{optimal.1} can be reformulated as
    \begin{align} \label{optimal.2}
        &\max\limits_{\mathbf{Q}} \quad\frac{1}{N}\sum\limits_{n=1}^N \log_2\left[1+\frac{P}{\sigma^2}{\left| \frac{A}{(d_{U\!G}[n])^{\kappa/2}} +\frac{B}{d_{U\!R}[n]}\right|^2} \right]\\
        &~~\textrm{s.t.} \quad  ~(1), \nonumber
    \end{align}
    where $A=\sqrt{\rho}|\tilde{h}|$, and $B=\sqrt{\rho}\sum_{i=1}^M |h_{RG,i}|$. 
    \subsection{Optimization of $\mathbf{Q}$}
    Problem \eqref{optimal.2} is still non-convex with respect to the UAV trajectory variables $\mathbf{Q}$. To tackle the non-convexity of problem \eqref{optimal.2}, we introduce slack variables
    $\mathbf{u}=\left\{ u[n]\right\} _{n=1}^{N}$ and $\mathbf{v}=\left\{ v[n]\right\} _{n=1}^{N}$, and consider the following problem:
    \begin{subequations}\label{slack.1}
    \begin{align}
        &\max\limits_{\mathbf{Q}, \mathbf{u}, \mathbf{v}} \quad\frac{1}{N}\sum\limits_{n=1}^N \log_{2}\left[1+\gamma_0\left|\frac{A}{(u[n])^{\kappa/2}} +\frac{B}{v[n]}\right|^{2}\right]\label{slack.1.a}\\
        &~~\textrm{s.t.} \quad~~  d_{U\!G}[n] \leq u[n], \forall n \label{slack.1.b}\\
        &\quad\quad ~~~~~d_{U\!R}[n] \leq v[n], \forall n \label{slack.1.c}\\
        &\quad\quad ~~~~~(1), \nonumber
    \end{align}
    \end{subequations}
    where $\gamma_{0}=P/\sigma^{2}$. Note that constraints \eqref{slack.1.b} and \eqref{slack.1.c} must hold with equality at the optimal solution of problem \eqref{slack.1}, since otherwise $u[n]$ and $v[n]$ can be increased to reduce the objective value. Therefore, problems \eqref{optimal.2} and \eqref{slack.1} have the same optimal solution. We next consider solving problem \eqref{slack.1}. To this end, we introduce an important lemma as follows.
    \begin{lem}
    Given $K_{1}>0$, $K_{2}>0$ and $K_{3}>0$, the function $f(x,y)=\log_{2}\left(1+\frac{K_{1}}{x^{\kappa}}+\frac{K_{2}}{y^{2}}+\frac{K_{3}}{x^{\kappa/2}y}\right)$
    is convex with respect to $x>0$ and $y>0$.
    \end{lem}
    \begin{IEEEproof}
    See Appendix.
    \end{IEEEproof}
    With Lemma 1, We see that the term $R_{U\!G}^{\text{slack}}[n]=\log_{2}\left[1+\gamma_0\left|\frac{A}{(u[n])^{\kappa/2}} +\frac{B}{v[n]}\right|^{2}\right]$ in \eqref{slack.1.a} is jointly convex with respect to $v[n]$ and $u[n]$. Note that the first-order Taylor approximation of a convex function is a global under-estimator. Hence, the first-order Taylor expansions of $R_{U\!G}^{\text{slack}}[n]$, $u^2[n]$, $v^2[n]$ at the given points $\mathbf{u}_0=\left\{ u_0[n]\right\} _{n=1}^{N}$ and $\mathbf{v}_0=\left\{v_0[n]\right\} _{n=1}^{N}$ can be respectively expressed as 
    \begin{align}
    &\log_{2}\left[1+\gamma_{0}\left[\frac{A^2}{(u[n])^{\kappa}}+\frac{B^2}{(v[n])^{2}}+\frac{2AB}{(u[n])^{\kappa/2}(v[n])}\right]\right]\nonumber\\
    &\geq\log_{2}A_{0}[n]+\frac{B_0[n]}{A_{0}[n]\ln2}(u[n]-u_{0}[n])\nonumber\\
    & ~~~~~~~~~~~~~~~~+\frac{C_0[n]}{A_{0}[n]\ln2}(v[n]-v_{0}[n]),
    \end{align}
    \begin{equation}
        -u^2[n] \leq u_0^2[n] - 2u_0[n]u[n],
    \end{equation}
    \begin{equation}
        -v^2[n] \leq v_0^2[n] - 2v_0[n]v[n],
    \end{equation}
    where
    $$A_0[n]=1+\gamma_{0}\left[\frac{A^2}{(u_0[n])^{\kappa}}+\frac{B^2}{(v_0[n])^2}+\frac{2AB}{(u_0[n])^{\kappa/2}(v_0[n])}\right],$$
    $$B_0[n]=-\gamma_0\left[\frac{\kappa A^2}{(u_{0}[n])^{(\kappa+1)}}+\frac{\kappa AB}{(v_0[n])(u_0[n])^{(\kappa/2+1)}}\right],$$
    and $$C_0[n]=-\gamma_0\left[\frac{2B^2}{(v_{0}[n])^{3}}+\frac{2AB}{(u_0[n])^{\kappa/2}(v_{0}[n])^{2}}\right].$$
    As such, problem \eqref{slack.1} can be approximated as
    \begin{subequations}\label{optimal.3}
    \begin{align}
        &\max\limits_{\mathbf{Q}, \mathbf{u}, \mathbf{v}} \quad\frac{1}{N} \sum\limits_{n=1}^N \frac{B_{0}[n]}{A_{0}[n]\ln2}u[n] +\frac{C_{0}[n]}{A_{0}[n]\ln2}v[n]\label{optimal.3.a}\\
        &~~~\textrm{s.t.} \quad  ~~(d_{U\!G}[n])^2+u_0^2[n]-2u_0[n]u[n] \leq 0 ,\forall n,\label{optimal.3.b}\\
        &\quad  ~~~~~~~~(d_{U\!R}[n])^{2}+v_0^2[n]-2v_0[n]v[n] \leq 0 ,\forall n,\label{optimal.3.c}\\
        &\quad  ~~~~~~~~(1).\nonumber
    \end{align}
    \end{subequations}
    Problem \eqref{optimal.3} is a convex optimization problem, and thus can be solved efficiently by standard solvers, such as the CVX.
    \subsection{Overall Algorithm}
    According to the obtained solutions in the previous two subproblems,  the overall algorithm for solving problem (9) is summarized in Algorithm 1, where $\epsilon$ is used to control the accuracy of convergence. Following the results in \cite{zhang2019}, we can guarantee that the average achievable rate by solving problem \eqref{optimal.1} is non-decreasing over iterations. Besides, the complexity of the proposed algorithm is $\mathcal{O}(K_{\rm{ite}}N^{3.5})$, where $K_{\rm{ite}}$ indicates the total number of iterations.
    \begin{algorithm}[t] 
    \caption{Proposed algorithm for solving \eqref{optimal.1}} 
    \label{alg1} 
    \begin{algorithmic}[1] 
    \STATE \textbf{Initialization}: Set initial variables $(\mathbf{Q}_0, \boldsymbol{\Phi}_0, \mathbf{u}_0, \mathbf{v}_0)$ and iteration number $k=0$. Set $\Bar{R}_0$ by using \eqref{rate.2} with given $(\mathbf{Q}_0, \boldsymbol{\Phi}_0)$.
    \STATE \textbf{repate}
    \STATE \quad Set $k\gets k+1$;
    \STATE \quad Update $(\mathbf{Q}_k, \mathbf{u}_{k}, \mathbf{v}_{k})$ by solving problem \eqref{optimal.3};
    \STATE \quad With given $\mathbf{Q}_k$, update $\boldsymbol{\Phi}_k$ by using \eqref{Phi.1};
    \STATE \quad With given $(\mathbf{Q}_k, \boldsymbol{\Phi}_k)$, update $\Bar{R}_k$ by using \eqref{rate.2}.
    \STATE \textbf{until}:  $\frac{\Bar{R}_k - \Bar{R}_{k-1}}{\Bar{R}_k} \textless \epsilon$.
    \end{algorithmic} 
    \end{algorithm}
\section{Numerical Results}
\begin{figure}[t]
        \centering
        \includegraphics[width=2.7in]{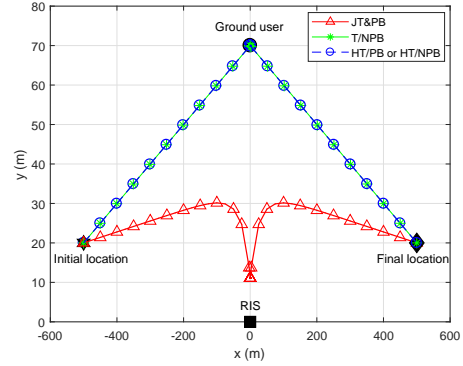}
        \caption{UAV trajectories by different algorithms when $T = 740$ $\rm{s}$}
        \label{figure_2}
    \end{figure}

    In this section, we provide simulation results for demonstrating the validity of the proposed joint UAV trajectory and RIS's passive beamforming optimization algorithm (denoted by JT\&PB). The following benchmark algorithms are used for comparison:
    \begin{itemize}
        \item UAV trajectory design without passive beamforming (referred to as T/NPB).
        \item Heuristic trajectory with passive beamforming (referred to as HT/PB).
        \item Heuristic trajectory without passive beamforming (referred to as HT/NPB).
    \end{itemize}
    Among these benchmarks, ``heuristic trajectory'' means that the UAV first flies directly to the ground user at $v_{max}$, then hovers above the user as long as possible, and finally flies to $\mathbf{q}_F$ at $v_{max}$ by the end of $T$. Naturally, the UAV trajectories of the HT/PB and HT/NPB algorithms are identical regardless of passive beamforming. Thus, we use the same marker to represent the trajectories of the two algorithms in Fig.~\ref{figure_2}. The remaining parameters are set as $\mathbf{q}_0 = [-500, 20]^T$ $\rm{m}$, $\mathbf{q}_F = [500, 20]^T$ $\rm{m}$, $\mathbf{w}_G=[0, 70]^T$ $\rm{m}$, $\mathbf{w}_R=[0, 0]^T$ $\rm{m}$, $z_U=80$ $\rm{m}$, $z_R=40$ $\rm{m}$, $v_{max}=25$ $\rm{m/s}$, $\delta_t=1$ $\rm{s}$, $M=90$, $\sigma^2=-80$ $\rm{dBm}$, $P=0.01$ $\rm{W}$, $d=\frac{\lambda}{2}$, $\alpha=2.8$, $\kappa=3.5$, $\beta=3$ $\rm{dB}$, $\rho=-20$ $\rm{dB}$, and $\epsilon = 10^{-4}$.
    
\begin{figure}[t]
        \centering
        \includegraphics[width=2.7in]{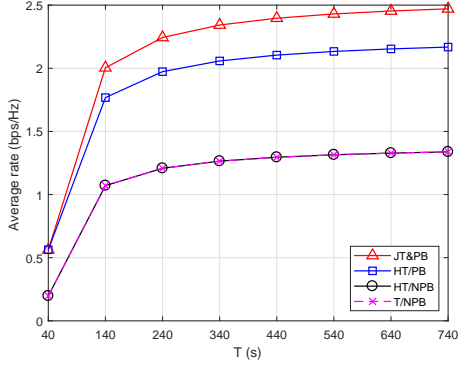}
        \caption{Average achievable rate performance by different algorithms verus $T$.}
        \label{figure_3}
    \end{figure}

    
    Fig.~\ref{figure_2} illustrates that the JT\&PB algorithm is different from all the benchmark algorithms from the perspective of UAV trajectory. When there is sufficient time for the UAV, the UAV will have the freedom to spend more time in hovering around a certain location to improve the average achievable rate. As such, we set a sufficiently large $T$, e.g., $T=740$ $\rm{s}$, to observe the variations of the trajectories optimized by all the algorithms. As shown in Fig.~\ref{figure_2}, we observe that when $T=740$ $\rm{s}$, the T/NPB, HT/PB, and HT/NPB algorithms find the same hovering location, namely, hovering above the ground user. Moreover, the more quickly the UAV approaches the hovering location, the higher the average achievable rate. Hence, the trajectories of the three benchmark algorithms are identical. Note that the trajectory of our JT\&PB algorithm is significantly different from those of the benchmark algorithms. The reason is that the JT\&PB algorithm can balance the channel gains between the U-G link and U-R-G link in each time slot to choose a trajectory, so as to achieve the best communication quality. This is also the reason why the trajectory of the JT\&PB algorithm is an arc path and its hovering location is different from those of the benchmark algorithms. With the help of Fig.~\ref{figure_3}, we also see that the JT\&PB algorithm achieves a considerable performance improvement compared with all the benchmark algorithms.

    Fig.~\ref{figure_3} shows the average achievable rates by all the considered algorithms versus $T$. Since with a large $T$ the UAV has sufficient time to transmit more information above its hovering location, the average achievable rates of all the benchmark schemes increase with $T$. Moreover, the average achievable rate of the proposed JT\&PB algorithm significantly exceeds that of the other algorithms. This demonstrates that the joint optimization of UAV trajectory and passive beamforming achieves a substantial gain, as compared with the counterpart approaches in which either UAV trajectory or passive beamforming is optimized. Thus, the ground user can enjoy the most benefit of the channel gains from both the UAV and the RIS, thereby obtaining the largest average achievable rate.
    
\section{Conclusions}
    In this letter, the RIS technique was applied to enhance the power of the received signal, thus further improving the achievable rate of the UAV-enabled communication system. To maximize the average achievable rate, we proposed a joint UAV trajectory and RIS's passive beamforming optimization algorithm for obtaining the high-quality suboptimal solution.
    Simulation results demonstrated that the assistance of the RIS is beneficial to substantially improve the communication quality of UAV-enabled networks.

\begin{appendix}
\section{}
\centerline{Proof of Lemma 1}
Lemma 1 is proved by the definition of convex function. First, the first-order partial derivatives of $f(x,y)$ with respect to $x$ and $y$ are given by
\begin{equation}
    \frac{\partial f}{\partial x}=\frac{-\kappa K_1x^{-\kappa-1}-(\kappa/2)K_3y^{-1}x^{-\kappa/2-1}}{\ln{2}(1+K_1x^{-\kappa}+K_2y^{-2}+K_3x^{-\kappa/2}y^{-1})},
\end{equation}
\begin{equation}
    \frac{\partial f}{\partial y}=\frac{-2K_2y^{-3}-K_3x^{-\kappa/2}y^{-2}}{\ln{2}(1+K_1x^{-\kappa}+K_2y^{-2}+K_3x^{-\kappa/2}y^{-1})}.
\end{equation}
Then, the second-order partial derivatives of $f(x,y)$ are given by
\begin{small}
\begin{align}
    &\frac{\partial^2 f}{\partial x^2}=\frac{\left[\kappa(\kappa+1)K_1x^{-\kappa-2}+(\kappa/2)(\kappa/2+1)K_3y^{-1}x^{-\kappa/2-2}\right]}{\ln{2}(1+K_1x^{-\kappa}+K_2y^{-2}+K_3x^{-\kappa/2}y^{-1})}\nonumber\\
    &\quad~~~~-\frac{(\kappa K_1x^{-\kappa-1}+(\kappa/2)K_3y^{-1}x^{-\kappa/2-1})^2}{\ln{2}(1+K_1x^{-\kappa}+K_2y^{-2}+K_3x^{-\kappa/2}y^{-1})^2},\\
    &\frac{\partial^2 f}{\partial y^2}=\frac{\left(6K_2y^{-4}+2K_3x^{-\kappa/2}y^{-3}\right)}{\ln{2}(1+K_1x^{-\kappa}+K_2y^{-2}+K_3x^{-\kappa/2}y^{-1})}\nonumber\\
    &\quad~~~~-\frac{(2K_2y^{-3}+K_3x^{-\kappa/2}y^{-2})^2}{\ln{2}(1+K_1x^{-\kappa}+K_2y^{-2}+K_3x^{-\kappa/2}y^{-1})^2},\\
    &\frac{\partial^2 f}{\partial x\partial y}=\frac{\kappa/2K_3x^{-\kappa/2-1}y^{-2}}{\ln{2}(1+K_1x^{-\kappa}+K_2y^{-2}+K_3x^{-\kappa/2}y^{-1})}\nonumber\\
    &-\frac{(\kappa K_1x^{-\kappa-1}\!+\!(\kappa/2)K_3y^{-1}x^{-\kappa/2-1})(2K_2y^{-3}\!+\!K_3x^{-\kappa/2}y^{-2})}{\ln{2}(1\!+\!K_1x^{-\kappa}\!+\!K_2y^{-2}\!+\!K_3x^{-\kappa/2}y^{-1})^2}.
\end{align}
\end{small}Therefore, the Hessian of $f(x,y)$ is
\begin{equation}
    \nabla^2f=\left[
    \begin{array}{cc}
         \frac{\partial^2 f}{\partial x^2} & \frac{\partial^2 f}{\partial x\partial y} \\
         \frac{\partial^2 f}{\partial y \partial x} & \frac{\partial^2 f}{\partial y^2}
    \end{array}
    \right].
\end{equation}
Since $\frac{\partial^2 f}{\partial x^2}>0$ and $\frac{\partial^2 f}{\partial x^2}\frac{\partial^2 f}{\partial y^2}-\frac{\partial^2 f}{\partial x\partial y}\frac{\partial^2 f}{\partial y \partial x}>0$, the Hessian matrix $\nabla^2f$ is positive definite. Thus, $f(x,y)$ is a convex function.
\end{appendix}

\section{Supplementary Information}
The proof of Lemma 1 are detailed as follows. Let $\eta = \ln{2}(1+K_1x^{-\kappa}+K_2y^{-2}+K_3x^{-\frac{\kappa}{2}}y^{-1})^2$. Then,
\begin{align}\label{proof.1}
    &\eta\frac{\partial^2 f}{\partial x^2}\nonumber\\ &=\kappa(\kappa\!+\!1)K_1x^{-\kappa-2}\!\left(1\!+\!K_1x^{-\kappa}\!+\!K_2y^{-2}\!+\!K_3x^{-\frac{\kappa}{2}}y^{-1}\right)\nonumber\\
    &+\frac{\kappa}{2}(\frac{\kappa}{2}\!+\!1)K_3y^{-1}x^{-\frac{\kappa}{2}-2}\!\left(1\!+\!K_1x^{-\kappa}\!+\!K_2y^{-2}\!+\!K_3x^{-\frac{\kappa}{2}}y^{-1}\right)\nonumber\\
    &-\left(\kappa K_1x^{-\kappa-1}+\frac{\kappa}{2}K_3y^{-1}x^{-\frac{\kappa}{2}-1}\right)^2\nonumber\\
    &=\kappa(\kappa\!+\!1)K_1x^{-\kappa-2}\!+\!\frac{\kappa}{2}\!\left(\frac{\kappa}{2}\!+\!1\right)K_3y^{-1}x^{-\frac{\kappa}{2}-2}\!+\!\kappa K_1^2x^{-2\kappa-2}\nonumber\\
    &+\!\left(\frac{\kappa^2}{4}\!+\!\frac{3\kappa}{2}\right)\!K_1K_3y^{-1}x^{-\frac{3\kappa}{2}-2}+\kappa(\kappa+1)K_1K_2x^{-\kappa-2}y^{-2}\nonumber\\
    &+\frac{\kappa}{2}\left(\frac{\kappa}{2}+1\right)K_2K_3y^{-3}x^{-\frac{\kappa}{2}-2}+\frac{\kappa}{2}K_3^2y^{-2}x^{-\kappa-2},
\end{align}
\begin{align}\label{proof.2}
    &\eta\frac{\partial^2 f}{\partial y^2}\nonumber\\
    &=6K_2y^{-4}\left(1\!+\!K_1x^{-\kappa}\!+\!K_2y^{-2}\!+\!K_3x^{-\frac{\kappa}{2}}y^{-1}\right)\nonumber\\
    &\quad+2K_3x^{-\frac{\kappa}{2}}y^{-3}\left(1\!+\!K_1x^{-\kappa}\!+\!K_2y^{-2}\!+\!K_3x^{-\frac{\kappa}{2}}y^{-1}\right)\nonumber\\
    &\quad-\left(2K_2y^{-3}\!+\!K_3x^{-\frac{\kappa}{2}}y^{-2}\right)^2\nonumber\\
    &= 6K_2y^{-4}\!+\!2K_3x^{-\frac{\kappa}{2}}y^{-3}\!+\!6K_1K_2x^{-\kappa}y^{-4}\!+\!2K_2^2y^{-6}\nonumber\\
    &\quad+4K_2K_3x^{-\frac{\kappa}{2}}y^{-5}+K_3^2x^{-\kappa}y^{-4}+2K_1K_3x^{-\frac{3\kappa}{2}}y^{-3},
\end{align}
\begin{align}\label{proof.3}
    &\eta\frac{\partial^2 f}{\partial x\partial y}\nonumber\\
    &=\frac{\kappa}{2}K_3x^{-\frac{\kappa}{2}-1}y^{-2}\!\left(1\!+\!K_1x^{-\kappa}\!+\!K_2y^{-2}\!+\!K_3x^{-\frac{\kappa}{2}}y^{-1}\right)\nonumber\\
    &-\left(\kappa K_1x^{-\kappa-1}\!+\!\frac{\kappa}{2}K_3y^{-1}x^{-\frac{\kappa}{2}-1}\right)\left(2K_2y^{-3}\!+\!K_3x^{-\frac{\kappa}{2}}y^{-2}\right)\nonumber\\
    &=\frac{\kappa}{2}K_3x^{-\frac{\kappa}{2}-1}y^{-2}-\frac{\kappa}{2}K_1K_3x^{-\frac{3\kappa}{2}-1}y^{-2}\nonumber\\
    &~~~-\frac{\kappa}{2}K_2K_3x^{-\frac{\kappa}{2}-1}y^{-4}-2\kappa K_1K_2x^{-\kappa-1}y^{-3},
\end{align}
\begin{align}\label{proof.4}
    &\eta^2\frac{\partial^2 f}{\partial x\partial y}\frac{\partial^2 f}{\partial y \partial x}\nonumber\\
    &=-\frac{\kappa^2}{2}K_2K_3^2x^{-\kappa-2}y^{-6}-2\kappa^2K_1K_2K_3x^{-\frac{3\kappa}{2}-2}\nonumber\\
    &\quad-\frac{\kappa^2}{2}K_1K_3^2x^{-2\kappa-2}y^{-4}\nonumber\\
    &\textcolor[rgb]{0,0,1}{\quad+\frac{\kappa^2}{4}K_3^2x^{-\kappa-2}y^{-4}+\frac{\kappa^2}{4}K_1^2K_3^2x^{-3\kappa-2}y^{-4}}\nonumber\\
    &\textcolor[rgb]{0,0,1}{\quad+\frac{\kappa^2}{4}K_2^2K_3^2x^{-\kappa-2}y^{-8}+4\kappa^2K_1^2K_2^2x^{-2\kappa-2}y^{-6}}\nonumber\\
    &\textcolor[rgb]{0,0,1}{\quad+2\kappa^2K_1K_2^2K_3x^{-\frac{3\kappa}{2}-2}y^{-7}+\frac{\kappa^2}{2}K_1K_2K_3^2x^{-2\kappa-2}y^{-6}}\nonumber\\
    &\textcolor[rgb]{0,0,1}{\quad+2\kappa^2K_1^2K_2K_3x^{-\frac{5\kappa}{2}-2}y^{-5},}
\end{align}
and
\begin{align}\label{proof.5}
    &\eta^2\frac{\partial^2 f}{\partial x^2}\frac{\partial^2 f}{\partial y^2}\nonumber\\
    &=6\kappa\left(\kappa\!+\!1\right)\!K_1\!K_2x^{-\kappa-2}y^{-4}\!+\!3\kappa\left(\frac{\kappa}{2}\!+\!1\right)\!K_2K_3x^{-\frac{\kappa}{2}-2}y^{-5}\!\nonumber\\
    &+6\left(\frac{\kappa^2}{4}+\frac{3\kappa}{2}\right)K_1K_2K_3x^{-\frac{3\kappa}{2}-2}y^{-5}+6\kappa K_1^2K_2x^{-2\kappa-2}y^{-4}\nonumber\\
    &+4\kappa(\kappa+1)K_1K_2K_3x^{-\frac{3\kappa}{2}-2}y^{-5}+2\kappa K_1^2K_3x^{-\frac{5\kappa}{2}-2}y^{-3}\nonumber\\
    &+2\kappa(\kappa+1)K_1K_3x^{-\frac{3\kappa}{2}-2}y^{-3}\nonumber+6\kappa K_1^3K_2x^{-3\kappa-2}y^{-4}\\
    &+2\kappa(\kappa+1)K_1K_2K_3x^{-\frac{3\kappa}{2}-2}y^{-5}+\kappa K_2^2K_3^2x^{-\kappa-2}y^{-8}\nonumber\\
    &+6\kappa(\kappa+1)k_1^2K_2x^{-2\kappa-2}y^{-4}+3\kappa K_1K_2K_3^2x^{-2\kappa-2}y^{-6}\nonumber\\
    &+3\kappa\left(\frac{\kappa}{2}+1\right)K_1K_2^2K_3x^{-\frac{3\kappa}{2}-2}y^{-7}+\kappa K_3^3x^{-\frac{3\kappa}{2}-2}y^{-5}\nonumber\\
    &+6\left(\frac{\kappa^2}{4}+\frac{3\kappa}{2}+\frac{2\kappa}{3}\right)K_1^2K_2K_3x^{-\frac{5\kappa}{2}-2}y^{-5}+\frac{\kappa}{2}K_3^4x^{-2\kappa-2}y^{-6}\nonumber\\
    &+\kappa\left(\frac{\kappa}{2}+1\right)K_1K_2K_3^2x^{-2\kappa-2}y^{-6}+\kappa K_1K_3^3x^{-\frac{5\kappa}{2}-2}y^{-5}\nonumber\\
    &+\kappa\left(\frac{\kappa}{2}+1\right)K_2^2K_3x^{-\frac{\kappa}{2}-2}y^{-7}+2\kappa K_1^2K_2^2x^{-2\kappa-2}y^{-6}\nonumber\\
    &+\kappa\left(\frac{\kappa}{2}+1\right)K_2^3K_3x^{-\frac{\kappa}{2}-2}y^{-9}+\!2\kappa(\kappa+1)K_1^2K_3x^{-\frac{5\kappa}{2}-2}y^{-3}\nonumber\\
    &+2\kappa\left(\frac{\kappa}{2}\!+\!1\right)\!K_2K_3^2x^{-\kappa-2}y^{-6}+2\kappa(\kappa+1)K_1K_2^2x^{-\kappa-2}y^{-6}\nonumber\\
    &+\frac{\kappa}{2}\left(\frac{\kappa}{2}\!+\!1\right)\!K_3^3x^{-\frac{3\kappa}{2}-2}y^{-5}\!+\!\left(\frac{\kappa^2}{4}\!+\!\frac{3\kappa}{2}\right)\!K_1\!K_3^3x^{-\frac{5\kappa}{2}-2}y^{-5}\nonumber\\
    &+3\kappa\left(\frac{\kappa}{2}+1\right)K_1K_2K_3x^{-\frac{\kappa}{2}-2}y^{-7}+3\kappa K_2K_3^2x^{-\kappa-2}y^{-6}\nonumber\\
    &+3\kappa\left(\frac{\kappa}{2}+1\right)K_1K_2K_3x^{-\frac{3\kappa}{2}-2}y^{-5}+2\kappa K_2K_3^3x^{-\frac{3\kappa}{2}-2}y^{-7}\nonumber\\
    &+\kappa\left(\frac{\kappa}{2}+1\right)K_1K_3^2x^{-2\kappa-2}y^{-4}+\kappa\left(\frac{\kappa}{2}+1\right)K_2K_3^2x^{-\kappa-2}y^{-6}\nonumber\\
    &+4\!\left(\frac{\kappa^2}{4}\!+\!\frac{3\kappa}{2}\right)\!K_1\!K_2K_3^2x^{-2\kappa-2}y^{-6}+2\kappa K_1^3K_3x^{-\frac{7\kappa}{2}-2}y^{-3}\nonumber\\
    &+\kappa\left(\frac{3\kappa}{2}+4\right)K_1K_3^2x^{-2\kappa-2}y^{-4}+2\kappa(\kappa+1)K_1K_2^3x^{-\kappa-2}y^{-8}\nonumber\\
    &+\frac{\kappa}{2}\left(\frac{\kappa}{2}\!+\!1\right)\!K_2K_3^3x^{-\frac{3\kappa}{2}-2}y^{-7}+6\kappa(\kappa+1)K_1K_2^2x^{-\kappa-2}y^{-6}\nonumber\\
    &+\!2\!\left(\frac{\kappa^2}{4}\!+\!\frac{3\kappa}{2}\right)\!K_1K_2^2K_3x^{-\frac{3\kappa}{2}-2}y^{-7}\!+\!\kappa K_1^2K_3^2x^{-3\kappa-2}y^{-4}\nonumber\\
    &\textcolor[rgb]{0,0,1}{+\left(\frac{\kappa^2}{2}+\kappa\right)K_3^2x^{-\kappa-2}y^{-4}+\left(\frac{\kappa^2}{2}+3\kappa\right)K_1^2K_3^2x^{-3\kappa-2}y^{-4}}\nonumber\\
    &\textcolor[rgb]{0,0,1}{+\left(\kappa^2+2\kappa\right)K_2^2K_3^2x^{-\kappa-2}y^{-8}+(6\kappa^2+6\kappa)K_1^2K_2^2x^{-2\kappa-2}y^{-6}}\nonumber\\
    &\textcolor[rgb]{0,0,1}{+(4\kappa^2+4\kappa)K_1K_2^2K_3x^{-\frac{3\kappa}{2}-2}y^{-7}+(\kappa^2+\kappa)K_1K_2K_3^2x^{-2\kappa-2}y^{-6}}\nonumber\\
    &\textcolor[rgb]{0,0,1}{+(2\kappa^2+2\kappa)K_1^2K_2K_3x^{-\frac{5\kappa}{2}-2}y^{-5}.}
\end{align}
Sine $K_{1}>0$, $K_{2}>0$, $K_{3}>0$, $x>0$, $y>0$, and the path loss exponent $\kappa>0$, it is observed that the three terms of $\eta\frac{\partial^2 f}{\partial x^2}$,  $\eta\frac{\partial^2 f}{\partial y^2}$, and $\eta^2\frac{\partial^2 f}{\partial x^2}\frac{\partial^2 f}{\partial y^2}$ are all positive, i.e., $\eta\frac{\partial^2 f}{\partial x^2}>0$,  $\eta\frac{\partial^2 f}{\partial y^2}>0$, and $\eta^2\frac{\partial^2 f}{\partial x^2}\frac{\partial^2 f}{\partial y^2}>0$. Since $\eta>0$, we have $\frac{\partial^2 f}{\partial x^2}>0$ and $\frac{\partial^2 f}{\partial y^2}>0$. Note that the blue part in \eqref{proof.4} is smaller than the blue part in \eqref{proof.5}, and that the remaining three terms in \eqref{proof.4} are all negative. Hence,  $\frac{\partial^2 f}{\partial x^2}\frac{\partial^2 f}{\partial y^2}-\frac{\partial^2 f}{\partial x\partial y}\frac{\partial^2 f}{\partial y \partial x}>0$. Since $\frac{\partial^2 f}{\partial x^2}>0$ and $\frac{\partial^2 f}{\partial x^2}\frac{\partial^2 f}{\partial y^2}-\frac{\partial^2 f}{\partial x\partial y}\frac{\partial^2 f}{\partial y \partial x}>0$, the Hessian matrix $\nabla^2f$ is positive definite. 

\end{document}